\newcommand{\abs}[1]{\left\vert#1\right\vert}

\newcommand{\ket}[1]{\left\vert #1\right\rangle}

\newcommand{\bra}[1]{\left\langle #1\right\vert}

\newcommand{\brkt}[2]{\left\langle #1 \vert #2\right\rangle}

\newcommand{\braket}[3]{\left\langle #1 \right\vert #2\left\vert #3\right\rangle}

\newcommand{\D}{\displaystyle}

\newcommand{\SSt}{\scriptscriptstyle}

\documentclass[preprint,aps,prc,showpacs,amsmath,amssymb]{revtex4-1}
\usepackage{times}          
\usepackage{graphicx}
\usepackage{bm}
\usepackage{amsmath}
\usepackage{amsxtra}
\usepackage{amstext}
\usepackage{amssymb}
\usepackage{latexsym}
\usepackage{dsfont}

\begin{document}
\title{{\large  $\mathbf{\Omega^{-}}$, $\mathbf{\Xi^{*\,-}}$, $\mathbf{\Sigma^{*\,-}}$ and $\mathbf{\Delta^{-}}$ decuplet baryon magnetic moments}}
\author{Milton Dean Slaughter}
\address{Department of Physics, Florida International University, Miami, Florida 33199, USA}
\email{slaughtm@FIU.Edu,Slaughts@PhysicsResearch.Net}

\begin{abstract}
The properties of the ground-state $U$-spin $=\frac{3}{2}$ baryon decuplet magnetic moments $\Delta^{-}$, $\Xi^{*\,-}$, $\Sigma^{*\,-}$ and $\Omega^{-}$ and their ground-state spin $\frac{1}{2}$ cousins $p$, $n$, $\Lambda$, $\Sigma^{+}$, $\Sigma^{0}$, $\Sigma^{-}$, $\Xi^{+}$, and $\Xi^{-}$ have been studied for many years with a modicum of success.  The magnetic moments of many are yet to be determined. Of the decuplet baryons, only the magnetic moment of the $\Omega^{-}$ has been accurately determined.  We calculate the magnetic moments of the \emph{physical} decuplet $U$-spin $=\frac{3}{2}$ quartet members without ascribing any specific form to their quark structure or intra-quark interactions.
\end{abstract}
\maketitle

\section{Introduction}

The properties of the ground-state $U$-spin $=\frac{3}{2}$ baryon decuplet magnetic moments $\Delta^{-}$, $\Xi^{*\,-}$, $\Sigma^{*\,-}$ and $\Omega^{-}$ along with their ground-state spin-$\frac{1}{2}$ cousins $p$, $n$, $\Lambda$, $\Sigma^{+}$, $\Sigma^{0}$, $\Sigma^{-}$, $\Xi^{+}$, and $\Xi^{-}$ have been studied for many years with a modicum of success.  Although the masses (pole or otherwise) and decay aspects and other physical observables of some of these particles have been ascertained, the magnetic moments of many are yet to be determined.  For the spin $=\frac{3}{2}$ baryon decuplet, the experimental situation is poor---from the Particle Data Group \cite{Amsler:2008zzb}, only the magnetic moment of the $\Omega^{-}$ has been accurately determined.  The reasons for this paucity of data for the decuplet particle members are the very short lifetimes owing to available strong interaction decay channels and the existence of nearby particles with quantum numbers that allow for configuration mixing.  The $\Omega^{-}$ is an exception in that it is composed of three valence $s$ quarks that make its lifetime substantially longer (weak interaction decay) than any of its decuplet partners, which have many more decay channels available.

A number of theoretical models have been put forth over the past few decades. In addition to the simplest $SU(3)$ model, seminal ones are the $SU(6)$ models put forth by Beg \emph{et al.} \cite{Beg:1964nm} and Gerasimov \cite{Gerasimov:1966}.  An excellent source of information on the aforementioned topics, references, and other seminal models is the book by Lichtenberg \cite{Lichtenberg:1978pc}.  Typically, these models invoke the additivity hypothesis where a hadron magnetic moment is given by the sum of its constituent quark magnetic moments.  More recently,
a number of theoretical and computational investigations involving the magnetic moments of the $\Omega^{-}$ and the $\Delta^{-}$ and lattice quantum chromodynamics (QCD) (quenched and unquenched, unphysical pion mass) techniques have been used with apparent progress and show promise \cite{Ramalho:2009gk,Boinepalli:2009sq,Aubin:2009qp}.  A review that focuses on some theoretical and experimental approaches to the study of specific processes involving the $\Delta(1232)$ can be found in Ref.~\cite{Pascalutsa:2006up}.

In this article the infinite momentum frame---in conjunction with the fact that the four-vector electromagnetic current $j^{\mu}_{em}$ obeys the equal time commutator $\left[V_{K^{0}}, j^{\mu}_{em}\right]=0$ even in the presence of symmetry breaking---is used to calculate the magnetic moments of the \emph{physical} decuplet $U$-spin $=\frac{3}{2}$ quartet members without ascribing any specific form to their quark structure or intra-quark interactions \cite{Oneda:1970ny,Oneda:1985wf,Slaughter:1988hx,Oneda:1989ik,alfaro63}.

\section{ETCRs in the Infinite Momentum Frame}
In this article all equal-time commutation relations (ETCRs) involve at most one current density, thus, problems associated with Schwinger terms are avoided.   ETCRs involve the vector and axial-vector charge generators (the
 $V_{\alpha }$ and $A_{\alpha }$ $\{\alpha
=\pi ,K,D,F,B,\ldots .\}$) of the symmetry groups of QCD and can be derived from a simple model $%
{\mathcal {L}}_{QCD}$ including quark mass terms. However, they
are valid even though these symmetries are broken
\cite{Oneda:1970ny,Oneda:1985wf,Slaughter:1988hx,Oneda:1989ik,alfaro63,gell-mann63,adler,Weisberger} and {\em even when the Lagrangian is
not known or cannot be constructed}. \ Some examples (summation over the dummy index $k$ is understood)
of these ETCRs or constraint algebras are:  $%
\left[V_{i},V_{j}\right]=\left[A_{i},A_{j}\right]=if_{ijk}V_{k}$, and $\left[V_{i},A_{j}\right]=if_{ijk}A_{k}$%
.

Mathematically, we have the following:  We introduce quark spinor
fields ${q_{\alpha}}^{i}$ each with mass $m_{i}$ where
$i=1,\cdots,N$ in the flavor $SU_{F}(N)$ group, and $\alpha=1,2,3$
are the color indices corresponding to the color $SU(3)$ group.
Then, we have (suppressing color indices):
\begin{equation}\label{eq01}
{{V_{a}}^{\mu}}(x)=i\bar{q}^{i}(x){(\lambda_{a}/2)}_{ij}
\gamma^{\mu}q^{j}(x)\equiv i\bar{q}{(\lambda_{a}/2)} \gamma^{\mu}q ,
\end{equation}
\begin{equation}\label{eq02}
{V}_{a}(t)=\int d^{3}x \; \mathbf{:}\,q^{\dag}(x){(\lambda_{a}/2)} q(x)\, \mathbf{:} \, ,
\end{equation}
\begin{equation}\label{eq03} {\partial}_{\mu}
{{V_{a}}^{\mu}}(x)=i{\bar{q}}^{i} (m_{i}-m_{j})
{(\lambda_{a}/2)}_{ij} q^{j} ,
\end{equation}
\begin{equation}\label{eq04}
{{A_{a}}^{\mu}}=i\bar{q}^{i}{(\lambda_{a}/2)}_{ij}
\gamma^{\mu}\gamma_{5}q^{j}\equiv i\bar{q}{(\lambda_{a}/2)}
\gamma^{\mu}\gamma_{5}q ,
\end{equation}
\begin{equation}\label{eq05}
{A}_{a}(t)=\int d^{3}x \; \mathbf{:}\, q^{\dag}(x){(\lambda_{a}/2)} \gamma_{5}q(x)\, \mathbf{:} \, ,
\end{equation}
\begin{equation}\label{eq06} {\partial}_{\mu}
{{A_{a}}^{\mu}}(x)=i{\bar{q}}^{i} (m_{i}+m_{j})
{(\lambda_{a}/2)}_{ij}\gamma_{5} q^{j} ,
\end{equation}
\begin{equation}\label{eq07}
[V_{a}(t),V_{b}(t)]=[A_{a}(t),A_{b}(t)]=if_{abc}V_{c}(t) ,
\end{equation}
and
\begin{equation}\label{eq08}
[V_{a}(t),A_{b}(t)]=if_{abc}A_{c}(t).
\end{equation}
In Eqs.~(\ref{eq01})--(\ref{eq08}), the $\lambda_{a}$,
$a=1,2,\cdots,N^2-1$, satisfy the {\it Lie algebra}
\begin{equation}\label{eq09}[(\lambda_{a}/2)(t),(\lambda_{b}/2)(t)]=if_{abc}(\lambda_{c}/2),\end{equation} where the $f_{abc}$ are
structure constants of the flavor group ${SU_{F}(N)}$ (summation over the dummy index $c$ is understood) and $\mathbf{:}\; \mathbf{:}$ denotes a normal product. We also have $\lambda_{0}=\sqrt{2/3} \;I$, $I$ is the identity, and ${{V_{0}}^{\mu}}(x)=i\bar{q}^{i}(x){(\lambda_{0}/2)}_{ij}
\gamma^{\mu}q^{j}(x)$ is the flavor singlet current.  Note that: (1) The vector charges $V_{a}(t)$ are not
conserved and are time dependent (except when $a=3,8,15, 24, 35$
and the $\lambda_{a}$ are diagonal corresponding to isospin,
strangeness, charm,...); (2) The axial-vector charges $A_{a}(t)$
are conserved only when {\em all quark masses vanish}; (3) The
ETCR algebras represented by Eqs.~(\ref{eq01})--(\ref{eq08}) {\em are
always valid even in the presence of symmetry breaking---the algebras are the same as those satisfied by the generators of unbroken $SU_{F}(N)$}; and (4)  In the flavor symmetry limit where $m_{u}=m_{d}=m_{s}=\ldots=m_{q}$, Eq.~(\ref{eq03}) shows that $ {\partial}_{\mu}
{{V_{a}}^{\mu}}=0$ and thus ${{V_{a}}^{\mu}}$ is conserved.

In terms of the axial-vector currents, ${A_{i}}^{\mu }$, partially conserved axial-vector current (PCAC)
is expressed by the equation, $\partial _{\mu }{A_{i}}^{\mu }(x)={m_{i}}%
^{2}f_{i}\phi _{i}(x)$, where $\phi _{i}(x)$ is the pseudoscalar field of
the particle $P_{i}$, $i=1,2,\ldots ,N^{2}-1$ in flavor symmetry $SU_{F}(N)$%
, and $f_{i}$ is defined by the expression (with our normalization) $\sqrt{(2%
{\pi }^{3})2p_{0}}\langle 0|{A_{i}}^{\mu }(0)|P_{i}(\vec{p})\rangle
=if_{i}p^{\mu }$ with $f_{i}=f_{i}(m_{i}^{2})$. In terms of physical
indices, we have ${{A_{\pi ^{\pm }}}^{\mu }}={{A_{1}}^{\mu }}\pm i{{A_{2}}^{\mu }}$%
, $\partial _{\mu }{{A_{\pi ^{+ }}}^{\mu }}(x)={m_{\pi ^{+}}}^{2}f_{\pi
^{+}}\phi _{\pi ^{+}}(x)$, $\partial _{\mu }{{A_{\pi ^{0 }}}^{\mu }}(x)=(1/%
\sqrt{2}){m_{\pi ^{0}}}^{2}f_{\pi ^{0}}\phi _{\pi ^{0}}(x)$, where $f_{\pi
^{+}}=$ is the pion decay constant.

\section{Infinite Momentum Frame Asymptotic ${\bf SU}_{F}{\bf
(N)}$ Symmetry}

A fundamental part of the dynamical concept of asymptotic
$SU_{F}(N)$ symmetry \cite{Oneda:1970ny,Oneda:1985wf,Slaughter:1988hx,Oneda:1989ik} is the behavior of the vector charge
$V_{\alpha }$ when acting on a physical state which has momentum
$\vec{k}$ ($|\vec{k}|\rightarrow \infty $), helicity $\lambda $,
and $SU_{F}(N)$ index $\alpha $: The physical annihilation
operator $a_{\alpha }(\vec{k},\lambda )$ of a physical
on-mass-shell hadron maintains its linearity (including asymptotic $%
SU_{F}(N)$ particle mixings) under flavor transformations
generated by the charge $V_{\alpha }$ but only in the limit
$|\vec{k}|\rightarrow \infty $. We note that the expression  $\vert \vec k\vert\rightarrow\infty$ is completely synonymous with the expression $\lim_{\vert\vec k\vert\rightarrow\infty}$. Thus, the $V_{\alpha }$ are
generators of asymptotic $SU_{F}(N)$ and have {\em {no}} \textquotedblleft
leakage\textquotedblright\ terms but {\em only} in this limit. See
Eq. (\ref{eq:matrixc}) below.

Consider the transformation (we suppress the time dependence) of the \emph{physical} annihilation operator $a_\alpha(%
\vec k,\lambda)$ under $SU_F(N)$ in \emph{broken symmetry}:
\begin{equation}
\left[V_i,a_\alpha(\vec k,\lambda)\right]=i\sum_{\beta}u_{i\alpha\beta} (\vec k%
,\lambda)a_\beta(\vec k, \lambda) +\delta u_{i\alpha \lambda}(\vec k) \;,
\label{eq:physop}
\end{equation}
where $\alpha$ and $\beta$ represent \emph{physical} $SU_F(N)$ indices, the coefficients $u_{i\alpha\beta} (\vec k%
,\lambda)=-(u_{i\beta\alpha} (\vec k%
,\lambda))^{*}$, $a_\beta(\vec k, \lambda)\stackrel{physical}{\vert \quad 0 \quad \rangle} =0$.
Although the term $\delta u_{i\alpha \lambda}(\vec k)=0$ \emph{in unbroken} $SU_F(N)$ symmetry, $\delta u_{i\alpha \lambda}(\vec k)\neq 0$ \emph{in broken} $SU_F(N)$ symmetry and is a function of the creation operators $a^{\dag}_\beta(\vec k, \lambda)$.

In \emph{exact unbroken} $SU_F(N)$, one writes instead
\begin{eqnarray}
\left[V_i,a_j(\vec k,\lambda)\right] & = & if_{ijk}a_k(\vec k,\lambda) \;for\;
j=1,2,\ldots,N^2-1\;,  \nonumber \\
& & \\
& = & 0\;for\;j=0 \;,  \nonumber  \label{eq:repop}
\end{eqnarray}
where $a_j(\vec k,\lambda)$ is a $SU_F(N)$ {\em {representation}}
annihilation operator, $a_j(\vec k, \lambda)\stackrel{representation}{\vert \qquad 0 \qquad \rangle} =0$, and its creation operator counterpart produces states that belong to irreducible representations of unbroken flavor $SU_F(N)$.  The {\em {dynamical}} assumption of asymptotic $%
SU_F(N)$ symmetry \cite{Oneda:1970ny,Oneda:1985wf,Slaughter:1988hx,Oneda:1989ik} then states that
\begin{equation}
\delta u_{i\alpha\lambda}(\vec k) \rightarrow (\vert \vec k\vert
)^{-(1+\epsilon)}, \; (\epsilon >0)\; \quad\mbox{when}\quad \vert \vec k \vert \rightarrow
\infty   \label{eq:deltau}
\end{equation}
implying that
\begin{equation}
\vert \alpha,\vec k, \lambda\rangle=\sum_j C_{\alpha j}\vert j, \vec k,
\lambda\rangle, \quad\mbox{when}\quad \vert \vec k \vert \rightarrow\infty.  \label{eq:matrixc}
\end{equation}

The orthogonal matrix $C_{\alpha j}$ depends on physical $SU_F(N)$ mixing
parameters, is defined only in the $\infty$-momentum frame, and can be {\sl {constrained directly by the
ETCRs without introducing an {\it ad hoc} mixing angle matrix}}. $\vert j, \vec k, \lambda \rangle $ is a $SU_F(N)$ representation
state whereas $\vert \alpha, \vec k,\lambda\rangle $ is a physical
state \cite{Oneda:1970ny,Oneda:1985wf,Slaughter:1988hx,Oneda:1989ik}.  All nonlinear terms vanish like ${\vert \vec k\vert}^{-(1+\epsilon)}$, ($%
\epsilon > 0$), as $\vert \vec k \vert \rightarrow\infty$, that is $\delta u_{i\alpha\lambda}(\vec k)\vert 0\rangle \sim \mathcal{O}(\vert \vec k\vert
)^{-(1+\epsilon)})\rightarrow 0$ as can be shown by applying Eq.~(\ref{eq:physop}) to the physical vacuum. It is in the $\infty
$-momentum frame where one finds that the {\em {physical}} annihilation operator
$a_{\alpha }(\vec{k},\lambda )$ is related {\em {linearly}} to the
{\em {representation}} annihilation operator
$a_{j}(\vec{k},\lambda )$ via the orthogonal mixing matrix
$C_{\alpha j}(\lambda )$. In contrast to the representation states $\vert j, \vec k,
\lambda\rangle$ that belong to irreducible representations, the states $\vert \alpha,\vec k, \lambda\rangle$ do not.  Rather, they are linear combinations of representation states plus non-linear corrective terms that are best calculated in a frame where mass differences are deemphasized such as the $\infty
$-momentum frame.   Thus, even in severely broken $SU_{F}(N)$ symmetry---such as $SU_{F}(4)$ or $SU_{F}(5)$---%
{\em {asymptotic}} $SU_{F}(N)$-symmetry methods are useful.  When flavor symmetry is exact, which Lorentz frame one uses to analyze current-algebraic sum rules does not matter and is a matter of taste, whereas, when one must deal with current-algebraic sum rules in broken symmetry, the choice of frame takes on paramount importance because one wishes to emphasize the calculation of leading order contributions while simultaneously simplifying the calculation of symmetry breaking corrections.  The $\infty
$-momentum frame is especially suited for broken symmetry calculations because mass differences are kinematically suppressed \cite{Oneda:1970ny,Oneda:1985wf,Slaughter:1988hx,Oneda:1989ik,alfaro63}.

The physical vector charge $V_{K^{0}}$ may be written as $V_{K^{0}}=V_{6}+iV_{7}$ and the physical electromagnetic current $j_{em}^{\mu}(0)$ may be written ($u$, $d$%
, $s$, $c$, $b$, $t$ quark system) as $j_{em}^{\mu}(0) =V_{3}^{\mu}(0)+(1/
\sqrt{3})V_{8}^{\mu}(0)-(2/3)^{1/2}V_{15}^{\mu}(0)+(2/5)^{1/2}V_{24}^{\mu}(0)- (3/5)^{1/2}V_{35}^{\mu}(0)
+(1\sqrt{3})V_{0}^{\mu}(0)$.  One may verify that the commutation relation $\left[V_{K^{0}},j^{\mu}_{em}(0)\right]=0$ holds (\emph{i.e.}, the electromagnetic current is a $U$-spin singlet).

\section{The Rarita-Schwinger Spinor}

For the on-mass shell $J^{P}=3/2^{+}$ ground state decuplet baryon B with mass $m_{B}$, the Lorentz- covariant and gauge-invariant electromagnetic current matrix element in momentum space where the four-momentum vectors $P\equiv p_{1}+p_{2},q\equiv p_{2}-p_{1}$ and $\lambda_{1}$ and $\lambda_{2}$ represent helicity is given by:
\begin{equation}
\left\langle B{(}p_{2},\lambda _{2})\right| j_{em}^{\mu}(0)\left|
B(p_{1}{,\lambda _{1})}\right\rangle ={\frac{e}{{(2\pi )^{3}}}}\sqrt{{%
\frac{{m}_{B}^{2}}{{E_{B}^{t}E_{B}^{s}}}}}\bar{u}^{\alpha}_{B}(p_{2},\lambda _{2})\left[ {%
\Gamma^{\mu }_{\alpha \beta}}\right] u^{\beta}_{B}\left( p_{1},\lambda _{1}\right),  \label{matrixeqn}
\end{equation}
\begin{equation}
\begin{array}{ccl}
\Gamma^{\mu }_{\alpha \beta} & = & g_{\alpha \beta}  \left\{ ( F_{1}^{B}(q^{2})+F_{2}^{B}(q^{2}) ) \gamma^{\mu
} - \D{\frac{P^{\mu}}{2 m_{B}}} F_{2}^{B}(q^{2})   \right\}\\
& + & \D{\frac{q_{\alpha}q_{\beta}}{(2 m_{B})^{2}}} \left\{ ( F_{3}^{B}(q^{2})+F_{4}^{B}(q^{2}) ) \gamma^{\mu
} - \frac{P^{\mu}}{2 m_{B}} F_{4}^{B}(q^{2})  \right\},%
\end{array}
\label{matrixcoveqn}
\end{equation}
\noindent where $e=+\sqrt{4\pi\alpha}$, $\alpha=$ the fine structure constant, the $F_{i}^{B}$ are the four $\gamma^{*}\Delta\Delta$ form factors and $\Gamma^{\mu }_{\alpha \beta}$ is written in a very useful form using the Gordon identities.  $Q_{B}=$ charge of baryon $B$ in units of $e$, $\mu_{B}$ is the magnetic moment (measured in nuclear magneton units $\mu_{N}=e/(2 m)$, $m=$ proton mass) of baryon $B$ and:
\begin{eqnarray}
  &&F_{1}^{B}(0)\,e  = Q_{B}\,,\\
  &&\mu_{B} = \left\{[F_{1}^{B}(0)+F_{2}^{B}(0)] (\frac{m}{m_{B}})\right\} \mu_{N} .
\end{eqnarray}

The baryon Rarita-Schwinger \cite{Rarita:1941mf} spinor $u_{B}^{\mu}(\nu_{B},\theta,\lambda)$ with helicity $\lambda$, three-momentum $\vec{p}$ with angle $\theta$ referred to the $\hat{z}$-axis, energy $E_{B}^{p}$, and velocity parameter $\nu_{B}=\sinh^{-1}({\abs{\vec p \,}}/m_B)$ is given by:
\begin{equation}
u_{B}^{\mu}(\nu_{B},\theta,\lambda)=
\sum_{m_1=-\frac{1}{2}}^{+\frac{1}{2}}\sum_{m_2=-1}^{+1}
\brkt{1/2, 1, 3/2}{m_1, m_2, \lambda}
u_{B}(\nu_{B},\theta,m_1)\epsilon_{B}^{\mu}(\nu_{B},\theta,m_2),
\label{rsspinoreqn}
\end{equation}
\begin{eqnarray}
u_{B}(\nu_{B},\theta,m_1)&=&
\left(
\begin{array}{l}
 \cosh(\frac{\nu_{B}}{2}) [ \cos(\frac{\theta}{2})\,\delta_{m_1,\,\frac{1}{2}} -\sin(\frac{\theta}{2})\,\delta_{m_1,\,-\frac{1}{2}} ]\\
 \cosh(\frac{\nu_{B}}{2}) [ \sin(\frac{\theta}{2})\,\delta_{m_1,\,\frac{1}{2}} +\cos(\frac{\theta}{2})\,\delta_{m_1,\,-\frac{1}{2}} ]  \\
 \sinh(\frac{\nu_{B}}{2}) [ \cos(\frac{\theta}{2})\,\delta_{m_1,\,\frac{1}{2}} +\sin(\frac{\theta}{2})\,\delta_{m_1,\,-\frac{1}{2}} ] \\
 \sinh(\frac{\nu_{B}}{2}) [ \sin(\frac{\theta}{2})\,\delta_{m_1,\,\frac{1}{2}} -\cos(\frac{\theta}{2})\,\delta_{m_1,\,-\frac{1}{2}} ]
 \end{array}
\right), \label{Bspinoreqn} \\ \nonumber
\end{eqnarray}
\begin{eqnarray}
\epsilon_{B}^{\mu}(\nu_{B},\theta,m_2)&=&
\left(
\begin{array}{l}
 \sinh(\nu_{B})\, \delta_{m_2,\,0}  \\
 -\frac{m_2}{\sqrt{2}}\cos(\theta)\,\delta_{\abs{m_2},\,1}+\cosh(\nu_{B}) \sin(\theta)\,\delta_{m_2,\,0}  \\
 -\frac{i}{\sqrt{2}} \,\delta_{\abs{m_2},\,1} \\
 \frac{m_2}{\sqrt{2}}\sin(\theta)\,\delta_{\abs{m_2},\,1}+\cosh(\nu_{B}) \cos(\theta)\,\delta_{m_2,\,0}
 \end{array}
\right). \label{polveceqn} \\ \nonumber
\end{eqnarray}
$\epsilon_{B}^{\mu}(\nu_{B},\theta,m_2)$ is the baryon polarization ($m_{2}$) four-vector, $u_{B}(\nu_{B},\theta,m_1)$ is a Dirac spinor with helicity index $m_{2}$, and $\brkt{1/2, 1, 3/2}{m_1, m_2, \lambda}$ is a Clebsh-Gordan coefficient where our conventions are those of Rose \cite{rose1957:etam}.  Physical states are normalized with $\brkt{\vec{p\,'}}{\vec{p}}=\delta^{3}%
(\vec{p\,'}\,-\vec{p})$ and Dirac spinors
are normalized by $\bar{u}^{(r)}(p)u^{(s)}(p)=\delta_{rs}$.  Our conventions for
Dirac matrices are $\left\{  \gamma^{\mu},\gamma^{\nu}\right\}
=2g^{\mu\nu}$ with $\gamma _{5}\equiv
i\gamma^{0}\gamma^{1}\gamma^{2}\gamma^{3}$, where $g^{\mu\nu}=$
Diag $(1,-1,-1,-1)$ \cite{Slaughter:2008zd}. The Ricci-Levi-Civita tensor is defined by
$\varepsilon _{0123}=-\varepsilon^{0123}=1=\varepsilon_{123}$.  As usual, we use natural units where $\hbar=c=1$.

Associated with baryon B are the four-momentum vectors $p_1$ (three-momentum $\vec{t}$ ($\vec{t}=t_z \hat{z}$), energy $E_{B}^{t}$) and $p_2$ (three-momentum $\vec{s}$ at angle $\theta$ ($0\leq\theta \leq \pi/2$) with the $\hat{z}$ axis, energy $E_{B}^{s}$) and we write:
\begin{eqnarray}
p_{1}^{\sigma}=t^{\sigma}=\left(
\begin{array}{c}
 m_{B}\cosh(\alpha_{B})\\
 0 \\
 0 \\
 m_{B}\sinh(\alpha_{B})
\end{array}
\right)=\left(
\begin{array}{c}
E_{B}^{t}\\
\vec{t}\\
\end{array}
\right)\label{momvecp1eqn}
,\\
p_{2}^{\sigma}=s^{\sigma}=\left(
\begin{array}{c}
 m_{B}\cosh(\beta_{B})\\
 m_{B}\sin(\theta) \sinh(\beta_{B})\\
 0 \\
 m_{B}\cos(\theta) \sinh(\beta_{B})
\end{array}
\right)=\left(
\begin{array}{c}
E_{B}^{s}\\
\vec{s}\\
\end{array}
\right).
\label{momvecp2eqn}
\end{eqnarray}

In Eqs.~(\ref{momvecp1eqn}) and (\ref{momvecp2eqn}), we take $s_{z}=r t_{z}$, where $r\, (\mbox{constant})\,\geq 1$.  In addition to obeying the Dirac equation---thus making the Gordon identities very useful--- the Rarita-Schwinger spinors satisfy the subsidiary conditions $\gamma_{\mu}u^{\mu}_{B}\left( p,\lambda\right)=p_{\mu}u^{\mu}_{B}\left( p,\lambda\right)=0$.
\section{The $\Omega^{-}$ and the $\Delta^{-}$ Magnetic Moment Relationship}
To obtain the relationship between the $\Omega^{-}$ and the $\Delta^{-}$ magnetic moments, we utilize the commutator  $\left[V_{K^{0}},j^{\mu}_{em}(0)\right]=0$ inserted between the baryon pairs ($\bra{\Xi^{*\,-}s^{\sigma}}$,$\ket{\Omega^{-}t^{\sigma}}$),
($\bra{\Sigma^{*\,-}s^{\sigma}}$,$\ket{\Xi^{*\,-}t^{\sigma}}$), and
($\bra{\Delta^{-}s^{\sigma}}$,$\ket{\Sigma^{*\,-}t^{\sigma}}$) in the infinite momentum frame where each baryon has $Q_{B}=-e$, helicity $+3/2$ and $t_{z}\rightarrow \infty$ and $s_{z}\rightarrow \infty$. The internal intermediate states saturating the commutator belong to the ground state decuplet baryons with helicity $+3/2$ which has the effect of restricting greatly the number of possible configuration mixing contributions coming from $56$ or spin $3/2$ members of $70$ excited states and other low-lying supermultiplets.  Given that caveat and noting that a vector charge does not change helicity or momentum , then with our normalization we have:
\begin{eqnarray}
\lefteqn{ \braket{\Xi^{*^-}s^{\sigma}}{V_{K^{0}}}{\Omega^{-}s^{\sigma}}   \braket{\Omega^{-}s^{\sigma}} {j_{em}^{\mu}} {\Omega^{-}t^{\sigma}} }  \nonumber \\
& & - \braket{\Xi^{*^-}s^{\sigma}}{j_{em}^{\mu}}{\Xi^{*^-}t^{\sigma}}   \braket{\Xi^{*^-}t^{\sigma}} {V_{K^{0}}} {\Omega^{-}t^{\sigma}} =0,\label{CRK1eqn}\\
\lefteqn{ \braket{\Sigma^{*\,-}s^{\sigma}}{V_{K^{0}}}{\Xi^{*^-}s^{\sigma}}   \braket{\Xi^{*^-}s^{\sigma}} {j_{em}^{\mu}} {\Xi^{*^-}t^{\sigma}} }  \nonumber \\
& & - \braket{\Sigma^{*^-}s^{\sigma}}{j_{em}^{\mu}}{\Sigma^{*^-}t^{\sigma}}   \braket{\Sigma^{*^-}t^{\sigma}} {V_{K^{0}}} {\Xi^{*^-}t^{\sigma}} =0,\label{CRK2eqn}\\
\lefteqn{ \braket{\Delta^{-}s^{\sigma}}{V_{K^{0}}}{\Sigma^{*^-}s^{\sigma}}   \braket{\Sigma^{*^-}s^{\sigma}} {j_{em}^{\mu}} {\Sigma^{*^-}t^{\sigma}} }  \nonumber \\
& & - \braket{\Delta^{-}s^{\sigma}}{j_{em}^{\mu}}{\Delta^{-}t^{\sigma}}   \braket{\Delta^{-}t^{\sigma}} {V_{K^{0}}} {\Sigma^{*^-}t^{\sigma}} =0. \label{CRK3eqn}
\end{eqnarray}

Now $\braket{\Xi^{*^-}s^{\sigma}}{V_{K^{0}}}{\Omega^{-}s^{\sigma}}=\braket{\Xi^{*^-}t^{\sigma}} {V_{K^{0}}} {\Omega^{-}t^{\sigma}}$, \emph{etc.} for each of the baryon pairs considered previously. Indeed, for exact flavor symmetry (or in broken symmetry in our model), the quantity $\braket{\Xi^{*^-}s^{\sigma}}{V_{K^{0}}}{\Omega^{-}s^{\sigma}}=\sqrt{3}$.  Equations~(\ref{CRK1eqn})--(\ref{CRK3eqn})  reduce to:
\begin{eqnarray}
\lefteqn{    \braket{\Omega^{-}s^{\sigma}} {j_{em}^{\mu}} {\Omega^{-}t^{\sigma}} }  \nonumber \\
& & - \braket{\Xi^{*^-}s^{\sigma}}{j_{em}^{\mu}}{\Xi^{*^-}t^{\sigma}}    =0,\label{CRK1eqnA}\\
\lefteqn{    \braket{\Xi^{*^-}s^{\sigma}} {j_{em}^{\mu}} {\Xi^{*^-}t^{\sigma}} }  \nonumber \\
& & - \braket{\Sigma^{*^-}s^{\sigma}}{j_{em}^{\mu}}{\Sigma^{*^-}t^{\sigma}}    =0,\label{CRK2eqnA}\\
\lefteqn{    \braket{\Sigma^{*^-}s^{\sigma}} {j_{em}^{\mu}} {\Sigma^{*^-}t^{\sigma}} }  \nonumber \\
& & - \braket{\Delta^{-}s^{\sigma}}{j_{em}^{\mu}}{\Delta^{-}t^{\sigma}}    =0. \label{CRK3eqnA}
\end{eqnarray}
Equations~(\ref{CRK1eqnA})--(\ref{CRK3eqnA}) then imply that
\begin{eqnarray}
\braket{\Delta^{-}s^{\sigma},\lambda}{j_{em}^{\mu}(0)}{\Delta^{-}t^{\sigma},\lambda}=\braket{\Omega^{-}s^{\sigma},\lambda} {j_{em}^{\mu}(0)} {\Omega^{-}t^{\sigma},\lambda},\nonumber \\
\mbox{where}\; t_{z}\rightarrow \infty, \;  s_{z}\rightarrow \infty,\;\mbox{and}\;\lambda=\mbox{helicity}\;=+3/2.
\label{CRMaineqn}
\end{eqnarray}

Although Eq.~(\ref{CRMaineqn}) is reminiscent of what one obtains in pure unbroken $SU_{F}(N)$ symmetry with a $U$-spin singlet electromagnetic current, it is now obtained in \emph{broken} symmetry.  With $r\, (\mbox{constant})\,\geq 1$ thus ensuring no helicity reversal, we now explicitly evaluate Eq.~(\ref{CRMaineqn}) with $\mu=0$ and $\theta=0$ which implies that $s_{x}=0$  (collinear case) using Eqs.~(\ref{matrixeqn})--(\ref{momvecp2eqn}).  We obtain:
\begin{eqnarray}
\lim_{\stackrel{t_{z}\rightarrow +\infty}{\SSt s_{z}\rightarrow \SSt +\infty}} \{  \frac{1}{2} \cosh\left[\frac{\alpha_{\Delta^{-}} -\beta_{\Delta^{-}} }{2}\right] (2 F_1^{\Delta^{-}}(q^{2}_{\Delta^{-}}) +F_2^{\Delta^{-}}(q^{2}_{\Delta^{-}}) -F_2^{\Delta^{-}}(q^{2}_{\Delta^{-}})  \cosh\left[\alpha_{\Delta^{-}} +\beta_{\Delta^{-}} \right] ) \}  \nonumber\\
=\lim_{\stackrel{t_{z}\rightarrow +\infty}{\SSt s_{z}\rightarrow \SSt +\infty}} \{  \frac{1}{2} \cosh\left[\frac{\alpha_{\Omega^{-}} -\beta_{\Omega^{-}} }{2}\right] (2 F_1^{\Omega^{-}}(q^{2}_{\Omega^{-}}) +F_2^{\Omega^{-}}(q^{2}_{\Omega^{-}}) -F_2^{\Omega^{-}}(q^{2}_{\Omega^{-}})  \cosh\left[\alpha_{\Omega^{-}} +\beta_{\Omega^{-}} \right] ) \}. \nonumber\\
\label{CRMain1eqn}
\end{eqnarray}
Taking the limits in Eq.~(\ref{CRMain1eqn}) with $s_{z}=r t_{z}$ [$r\, (\mbox{constant})\,\geq 1$ and $s_{x}= 0$] yields:
\begin{equation}\label{CRMain2eqn}
F_2^{\Delta^{-}}(q^{2}_{\Delta^{-}})=\frac{m^{2}_{\Delta^{-}}}{m^{2}_{\Omega^{-}}}\, F_2^{\Omega^{-}}(q^{2}_{\Omega^{-}}).
\end{equation}
In deriving  Eq.~(\ref{CRMain2eqn}), we utilized that, in general, even though $\abs{\vec{s}}\,\mbox{and}\,\abs{\vec{t}}\rightarrow +\infty$, $q^{2}_{B}$ is finite and $q^{2}_{B} = -\frac{(1-r)^2}{r}m^2_{B}-\frac{s_{x}^2}{r}\equiv -Q^{2}_{B}$.
\begin{eqnarray}
  {q^{2}_{B}}_{\mid s_{x}=0} &=& -\frac{(1-r)^2}{r}m^2_{B}\,,
  \label{qsqcollineareqn}
\end{eqnarray}
\begin{eqnarray}
  \cosh\left[\frac{\alpha_{B} -\beta_{B} }{2}\right] &\rightarrow& \frac{1+r}{2 \sqrt{r}}\,, \nonumber\\
  \cosh\left[\alpha_{B} +\beta_{B} \right] &\rightarrow&  \frac{2 \,r\, t_{z}^2}{m_{B}^2}\,,
\end{eqnarray}
where $B=\Delta^{-} \;\mbox{or}\;\Omega^{-}$.

Setting $r=1$ $\Rightarrow q^{2}_{\Delta^{-}}=q^{2}_{\Omega^{-}}=0$, we obtain:
\begin{eqnarray}\label{CRMain3eqn}
F_1^{\Delta^{-}}(0) &=& F_{1}^{\Omega^{-}}(0)=-1\, ,\nonumber\\
F_2^{\Delta^{-}}(0)&=&\frac{m^{2}_{\Delta^{-}}}{m^{2}_{\Omega^{-}}}\, F_2^{\Omega^{-}}(0).
\end{eqnarray}
\begin{eqnarray}
  \label{CRMain4eqn}
  \mu_{\Delta^{-}} &=& \left[[F_{1}^{\Omega^{-}}(0)+
  \frac{m^{2}_{\Delta^{-}}}{m^{2}_{\Omega^{-}}}\, F_2^{\Omega^{-}}(0)
  ] (\frac{m}{m_{\Delta^{-}}})\right] \mu_{N}\nonumber \\
   &=&\left[[-1+
  \frac{m^{2}_{\Delta^{-}}}{m^{2}_{\Omega^{-}}}\, F_2^{\Omega^{-}}(0)
  ] (\frac{m}{m_{\Delta^{-}}})\right] \mu_{N}.
\end{eqnarray}

Experimentally \cite{Amsler:2008zzb}, $\mu_{\Omega^{-}}=(-2.02 \pm 0.05)\,\mu_{N}=\left[(-1+
  F_2^{\Omega^{-}}(0)
  ) (m/m_{\Omega^{-}})\right] \mu_{N}$ and $m_{\Omega^{-}}=1.6724\pm 0.0003\;GeV/c^{2}$, whereas the value of $m_{\Delta^{-}}$ (pole or Breit-Wigner) is not established.  We take $m_{\Delta^{-}}= 1.22\pm 0.01\;GeV/c^{2}$ and upon using Eqs.~(\ref{CRMain3eqn}) and (\ref{CRMain4eqn}), one obtains
$\mu_{\Delta^{-}}= (-1.83\pm 0.04)\,\mu_{N}$.  From Eqs.~(\ref{CRK1eqnA})--(\ref{CRMaineqn}), one can also calculate the magnetic moments of $\Xi^{*^-}$ and $\Sigma^{*^-}$.  We summarize these results in Table I.

\begin{table}[h]
\caption{\label{table1}$U$-spin $=\frac{3}{2}$  baryon decuplet magnetic moment $\mu$ in units of $\mu_{N}$.}
\begin{ruledtabular}
\begin{tabular}{@{}lcclcc@{}}
Baryon &  This research\;\footnotemark[1]  & Broken $SU_{F}(6)$\;\footnotemark[2]  & CST\;\footnotemark[3]  & Lattice QCD\;\footnotemark[4]  & Lattice QCD\;\footnotemark[5]  \\
       &    &          &      &       &  \\
\hline
$\Delta^{-}$\hphantom{00} & \hphantom{0}$-1.83\pm 0.04$ & \hphantom{0}$-2.92\pm 0.02$ & $-2.70 $  & $-1.697\pm 0.065$ &  $-1.85\pm 0.06$ \\
$\Sigma^{*\,-}$\hphantom{00} & \hphantom{0}$-1.89\pm 0.05$ & \hphantom{0}$-2.56\pm 0.01$ & $-2.44$   & $-1.697\pm 0.065$ & ---  \\
$\Xi^{*\,-}$\hphantom{0} & \hphantom{0}$-1.95\pm 0.05$ & \hphantom{0}$-2.20\pm 0.01$ & $-2.23$\hphantom{0}   & $-1.697\pm 0.065$ & --- \\
$\Omega^{-}$\hphantom{0} & \hphantom{0}$-2.02\pm 0.05$ & \hphantom{0}$-1.84\pm 0.02$ & $-2.02$  & $-1.697\pm 0.065$ & $-1.93\pm 0.08$  \\
\end{tabular}
\end{ruledtabular}
\footnotetext[1]{$\mu_{\Omega^{-}}$ is input.  $m_{\Delta^{-}}= 1.22\pm 0.01\,GeV/c^{2}$ is assumed.}

\footnotetext[2]{$\mu_{p}$, $\mu_{n}$, $\mu_{\Lambda}$ are inputs;  see Ref.~\cite{Lichtenberg:1978pc}}

\footnotetext[3]{Covariant spectator theory (CST);  see Ref.~\cite{Ramalho:2009gk}. $\mu_{p}$, $\mu_{n}$, $\mu_{\Omega^{-}}$ are inputs.}

\footnotetext[4]{Lattice result from Ref.~\cite{Boinepalli:2009sq}.  $\mu_{p}$, $\mu_{n}$, $\mu_{\Omega^{-}}=\mu_{\Delta^{-}}=-\mu_{\Delta^{+}}$ is assumed.}

\footnotetext[5]{Lattice result from Ref.~\cite{Aubin:2009qp}.}

\end{table}


\section{Conclusions}
We have calculated the magnetic moments of the ground state \emph{physical} decuplet $U$-spin $=\frac{3}{2}$ quartet members including that of $\Delta^{-}$ without ascribing any specific form to their quark structure or intraquark interactions or assuming $SU_F(N)$ broken or unbroken symmetry or assuming an effective lagrangian.  The Particle Data Group \cite{Amsler:2008zzb} value $\mu_{\Omega^{-}}=(-2.02 \pm 0.05) \,\mu_{N}$ was used as input.  Our results are compared to some extant lattice QCD results, results from $SU_F(N)$ models, and other theoretical models.  In particular, we obtain $\mu_{\Delta^{-}}= (-1.83 \pm 0.04) \,\mu_{N}$.

\newpage

\bibliographystyle{apsrev}

\end{document}